\documentclass[12pt]{iopart}
\usepackage{iopams}  

\usepackage{tikz}
\usetikzlibrary{quantikz}

\usepackage{graphicx}
\usepackage{dcolumn}
\usepackage{bm}
\usepackage{hyperref}

\usepackage{float} 
\usepackage{xcolor}
\usepackage[normalem]{ulem} 

\begin{document}

\title{Towards solving the BCS Hamiltonian gap in Near-Term Quantum Computers}

\author{Nahum S\'a}
\address{Centro Brasileiro de Pesquisas F\'isicas, Rua Dr. Xavier Sigaud 150, 22290-180 Rio de Janeiro, Brazil}
\ead{nahumsa@cbpf.br}
\author{Ivan S. Oliveira}
\address{Centro Brasileiro de Pesquisas F\'isicas, Rua Dr. Xavier Sigaud 150, 22290-180 Rio de Janeiro, Brazil}
\ead{ivan@cbpf.br}
\author{Itzhak Roditi}
\address{Centro Brasileiro de Pesquisas F\'isicas, Rua Dr. Xavier Sigaud 150, 22290-180 Rio de Janeiro, Brazil}
\ead{roditi@cbpf.br}

\date{\today}

\begin{abstract}
In this work, using a NISQ framework, we obtain the gap of a BCS Hamiltonian. This could lead to interesting implications for superconductivity research. For such task, we choose to use the Variational Quantum Deflation and analyze the hardware restrictions that are needed to find the energy spectra on current quantum hardware.

We also compare two different kinds of classical optimizers, Constrained Optimization BY Linear Approximations (COBYLA) and Simultaneous Perturbation Stochastic Approximation (SPSA), and study the effect of decoherence caused by the presence of noise when using simulations in real devices. We implement this method for a system with both 2 and 5 qubits. Furthermore, we show how to approximate the gap within one standard deviation, even with the presence of noise.
\end{abstract}

\maketitle


\section{Introduction}
\label{Sec: Introduction}

At the current time, we may consider that we are in the \textit{Noisy Intermediate-Scale Quantum} (NISQ) \cite{preskill2018NISQ} era of Quantum Computing which refers to devices with 50-100 qubits, without Quantum Error Correction. In other words, we are still far from devices with perfect qubits and quantum operations. Thus, in order to attain an acceptable performance, we need to use algorithms that are suited for those limitations. A very valuable class of algorithms appropriate to deal with such limitations, and that may be very useful in physics as well as in many other fields, e.g. machine learning, are the so-called Variational Quantum Algorithms (VQA).

Variational Quantum Algorithms is a class of hybrid classical-quantum algorithms that shows noise resilience due to the use of parametric quantum circuits \cite{reiner2019noisevqe}, thus being able to be implemented in NISQ devices. 
Variational Quantum Algorithms are universal \cite{biamonte2021universal} and have been proposed for various problems, such as Combinatorial Optimization \cite{farhi2014quantum}, Quantum Chemistry \cite{peruzzo2014VQE},  factoring \cite{anschuetz2019factoring}, Machine Learning \cite{schuld2020VQC} and compilation \cite{khatri2019compiling}.

One problem that can be dealt with by the use of such techniques and has the potential to lead to interesting outcomes is superconductivity. This is fairly well modeled using a pairing Hamiltonian.

BCS superconductivity is an appealing framework for testing VQA. Experimentally, the simulation of pairing Hamiltonians have been shown using a Nuclear Magnetic Resonance (NMR) setup \cite{Yang2006NMR}. However, due to intrinsic NMR experimental characteristics, there are several limitations for scalability which can be handled using VQAs.

In order to simulate Fermionic Hamiltonians, in digital quantum computers, one needs to map fermions to qubits. There are many mapping techniques used for quantum chemistry, for instance Jordan-Wigner \cite{Jordan1928}, and Bravyi-Kitaev Mapping \cite{bravyi2002fermionic}. One  of such
qubit mappings proposed by Wu et al. \cite{wu2002BCS2QUBIT} has been only implemented for NMR quantum computers, although it can be used for circuit-based quantum computers and more specifically NISQ devices using Variational Quantum Algorithms. This paper aims to use this mapping to solve the BCS Hamiltonian. On the other hand, one can also can simulate Fermionic Hamiltonians using analog quantum simulations \cite{Parsons2016,Bloch2012}.

In order to show that this mapping is suitable for circuit-based quantum computers, we will use the Fermion-to-Qubit mapping made by Wu et al. \cite{wu2002BCS2QUBIT} to solve the BCS problem with both $N=5$ and $N=2$ qubits. This also extends the result of \cite{Yang2006NMR} in a setting where it is not needed to use any simplification of the parameters of the Hamiltonian, leading to a path into solving this problem with an arbitrary number of cooper pairs which is limited only by quantum computer hardware. 

In order to demonstrate this, our goal is to find the energy spectra. One can use Variational Quantum Algorithms. We choose to use the Variational Quantum Deflation algorithm \cite{higgott2019VQD} which could lead to restrictions of the topology of quantum computers. Fortunately, those restrictions are met by real devices which could be used to solve the BCS Hamiltonian. However, this method can work with any algorithm able to find excited states on quantum computers, such as Quantum Subspace Expansion methods \cite{PhysRevX.8.011021, PhysRevResearch.1.033062}.

The paper is organized as follows: Section \ref{Sec: Methods} presents the BCS Hamiltonian, the qubit mapping used, explains the Variational Quantum Deflation algorithm, and the ansatz used. Section \ref{Sec:Results} shows the numerical simulation results for an ideal quantum computer and a noisy quantum computer assuming a Thermal Relaxation Error noise model. Lastly, in Section \ref{Sec: Conclusion}, we conclude the paper and present possible ideas for future research. In \ref{Appendix: Classical Optimizer} we explain the optimizers used in further detail. In \ref{Appendix: Swap Test} we further explain the methods that can be used to measure the overlap between two states using a quantum computer and the hardware restrictions.
 
\section{Methods}
\label{Sec: Methods}
First, we need to define the Fermionic Hamiltonian that we want to solve. The general Bardeen–Cooper–Schrieffer (BCS) \cite{bardeen1957BCS} Hamiltonian is given by: 

\begin{equation}
    \label{Eq: BCS Hamiltonian}
    H_{BCS} = \sum_{m=1}^N \frac{\epsilon_m}{2} ( n_m + n_{-m} ) \ + \sum_{m,k=1}^N V^{+}_{ml} c^\dagger_m c^\dagger_{-m} c_{-l}c_l
\end{equation} where $n_{\pm m} = c^\dagger_{\pm m} c_{\pm m}$ is the number operator, and the matrix elements $V^+_{ml} = \langle m, -m| V | l, -l \rangle$ are real and can be calculated efficiently for any given problem. 
For simplicity V can be treated as a constant which yields a good approximation to many superconductors \cite{waldram1976weaklysuperconductor}.

An important property that can be obtained from the Hamiltonian is the gap, which is defined as $2 \Delta_n \equiv E_{n,1} - E_{n,0}$ , $n$, being the number of Cooper pairs (see  \cite{wu2002BCS2QUBIT} for an NMR simulation experiment where the energy spectrum of the Hamiltonian is determined). For instance, when $n=0$, one has for the gap between the first excited state and the ground state, $2 \Delta_0 \equiv E_{1} - E_{0}$. And for $n=1$, one has the gap between the first and second excited states, $2 \Delta_1 \equiv E_{2} - E_{1}$. In the present work, we will focus both on the $n=1$ gap and the $n=0$ gap.


In order to solve this problem in a quantum computer, one needs to map the BCS Hamiltonian into a qubit Hamiltonian with a one-to-one mapping. In this paper we will use one of the mappings presented in Wu et al. \cite{wu2002BCS2QUBIT} which maps Fermions to qubits, given by the following representation:

\begin{equation}
    \label{Eq: Qubit Hamiltonian}
    H_Q = \sum_{m=1}^N \frac{\epsilon_m}{2} \sigma^Z_m + \frac{V}{2} \sum_{l>m=1}^N ( \sigma^x_m \sigma^x_l +  \sigma^y_m \sigma^y_l )
\end{equation}

One can find the gap by finding the energy spectra of the qubit Hamiltonian (Eq. \ref{Eq: Qubit Hamiltonian}) since the mapping assures that the qubit Hamiltonian has the same energy spectra as that of the BCS Hamiltonian. The goal of this paper is to find the energy spectra using a variational quantum algorithm called Variational Quantum Deflation \cite{higgott2019VQD}.

The Variational Quantum Deflation (VQD) algorithm is an extension of the Variational Quantum Eigensolver (VQE) \cite{peruzzo2014VQE} algorithm, allowing us to approximate excited states of a desired Hamiltonian, thus giving us access to their energy values. 

The original VQE algorithm only finds the ground energy of a given Hamiltonian by using the decomposition of such Hamiltonian into Pauli strings and minimizing the expected value of the Hamiltonian:

\begin{equation}
    \label{Eq: VQE Cost Function}
    E(\theta) = \langle \psi(\theta) | H | \psi(\theta) \rangle = \sum_i c_i \langle \psi(\theta) | P_i | \psi(\theta) \rangle 
\end{equation}

Where $P_i$ are Pauli strings that are of polynomial size. In order to measure this expectation value one needs to change from the computational basis into the X and Y basis, thus this approach has low depth and is suited for NISQ devices. By minimizing the cost function (Eq. \ref{Eq: VQE Cost Function}) it is possible to approximate the ground state of a Hamiltonian with high precision.

In order to get the energy of excited states, the cost function of the VQE needs to be modified by taking into account that all eigenstates of the Hamiltonian are orthogonal to each other. This is done by minimizing the overlap between all eigenstates discovered using the algorithm. Thus, to calculate the k-th excited state of a given Hamiltonian, $H$, the parameters $\lambda_k$ must be optimized for a parametrized state $| \psi (\lambda_k) \rangle$ for a given cost function. The cost function that needs to be minimized for the VQD is given by:

\begin{equation}
    \label{Eq: Cost Function VQD}
    F(\lambda_k) = \langle \psi(\lambda_k) | H | \psi(\lambda_k) \rangle + \sum_{i=0}^{k-1} \beta_i \big| \langle \psi(\lambda_k) | \psi(\lambda_i) \rangle \big|^2
\end{equation}

The first term is the expected value of the energy, which can be obtained measuring the Pauli decomposition of the Hamiltonian $H$ just like the VQE cost function. On the other hand, the second term is not as trivial, one needs to calculate the overlapping term between all eigenstates until the desired k-th eigenstate, this term ensures that all eigenstates are orthogonal to each other. 

It is possible to measure the overlap between two quantum states on quantum computers, this will be explained in more detail in \ref{Appendix: Swap Test}. In addition, those methods imposes hardware restrictions, which are also discussed in the appendix.

One of the downsides of the VQD algorithm is that the hyperparameter $\beta_i$ has to be chosen, but, according to  \cite{higgott2019VQD}, it suffices to choose $\beta_i > E_k - E_i $, where $E_k$ is the desired energy state and $E_i$ all energy states below the desired state, since this paper aims to find the energy spectra of the Hamiltonian, we choose $\beta_i > E_{\text{max}} - E_{\text{min}}$, where $E_{\text{max}}$ is the highest value of the energy and $E_{\text{min}}$ the minimum value of the energy. Another aspect to be careful with is that if the real ground state is not satisfactorily approximated, there may be an error propagation due to the initial state not being well-prepared.

In the following subsection, we will describe different kinds of variational ansatz and our choice of a Hardware-Efficient Ansatz in more detail.

\subsection{Variational Ansatz}
\label{Subsec: Variational Ansatz}
In order to build a variational ansatz for the Variational Quantum Algorithm, two conflicting goals should be met: The ansatz must be as general as possible in order to create states that would approximate all eigenstates of the Hamiltonian, but also needs to have a polynomial number of parameters in order to express our ansatz efficiently on a quantum computer.

In the literature there are two distinct paths for constructing the variational ansatz: 1) You can create an ansatz based on the problem's knowledge, in this case we can see as an example the UCCSD ansatz \cite{mcardle2020quantumchem}; 2) Since the main goal is to run the Variational Quantum Algorithm in real hardware one needs to take into account hardware restrictions on connectivity and the gate set and construct an ansatz that is suited for the Hardware that the algorithm will run on, this is commonly called Hardware Efficient Ansatz introduced by Kandala et al. \cite{kandala2017hardware}.

\begin{figure}[H]
    \centering
    \includegraphics[width=0.5\textwidth]{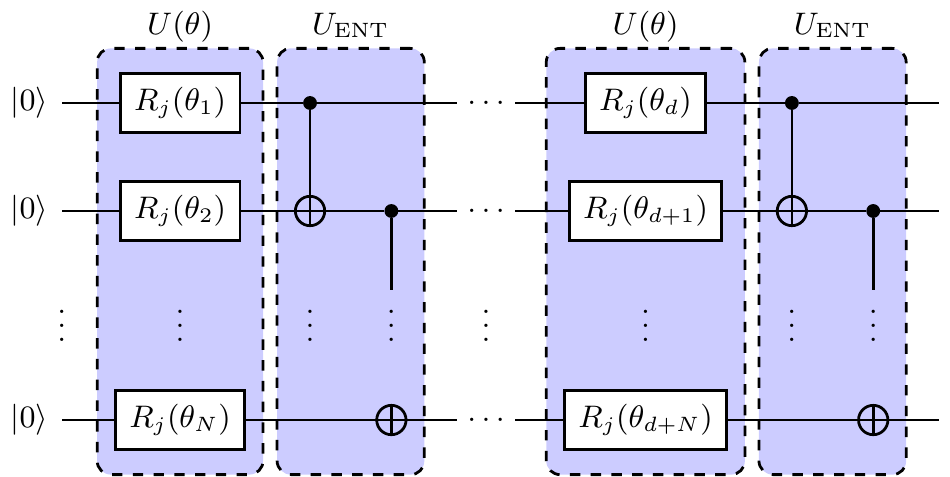}
    \caption{Hardware Efficient ansatz where the unitary $U_{\mathrm{ENT}}$ is constructed using the connection graph of the Quantum Hardware, and the unitary $U(\mathbb{\theta})$ is made of hardware-native parametrized rotations, in our case it is the combination of $R_Y$ and $R_Z$ rotations.}
    \label{fig: Hardware Efficient circuit}
\end{figure}

Our choice of ansatz for this paper is the {\em hardware efficient ansatz}. This kind of ansatz is depicted in FIGURE \ref{fig: Hardware Efficient circuit} and is composed by blocks of two kind of unitary transformations. The first block is made of one or more hardware-native parametrized one qubit gates ($R_j(\theta_i)$) and depicted on the figure as $U(\mathbb{\theta})$. 

The second block is constructed using two qubit gates that can generate entanglement, which are generally called entangling gates, the most common entangling gate is the CNOT, however there are other entangling gates such as iSWAP used on the Sycamore chip made by Google \cite{arute2019SupremacyGoogle}. Since entangling gates are the gates which have lower coherence, it is imperative to use them according to the hardware's connection graph in order to avoid using swap gates, which are costly for NISQ devices and can lead to non-local errors. 

Thus, the ansatz can be written as following:

\begin{equation}
    \label{Eq: Hardware Efficient Ansatz}
    \ket{\psi(\theta)} = U_{\text{ENT}} \  U(\theta) \ \dots  \  U_{\text{ENT}} \ U(\theta) \ket{\psi(0)}
\end{equation}

Depth, $d$, is defined as the number of both the single qubit rotations block and the entangling block. The dependency on the number of parameters regarding the depth is given by $2Nd$, where $N$ is the number of qubits, yielding a polynomial number of parameters, which ensures that we are working with an efficient ansatz construction.


\small

\section{Results}
\label{Sec:Results}

In order to demonstrate our proposed method, we choose to simulate it both on ideal and noisy quantum computers for the case of $N = 2$ and $N = 5$ qubits, because the system is solvable using direct diagonalization, and we can compare results between the result of direct diagonalization and the result from the quantum computer. Both simulations will take into account that we only take a finite amount of measurements (10,000 shots) from the quantum computer.

The Hamiltonian that we aim to solve is given by equation (\ref{Eq: Qubit Hamiltonian}) with $N=2,5$. An interesting parameter that we can obtain for this Hamiltonian, as mentioned before, is the gap: $2 \Delta_n = E_{n,1} - E_{n,0}$.

This section will be divided in two subsections, in which different parameters of the simulation will be considered. The first section \ref{Subsec: Ideal Simulation} will be the simulation performed by a perfect quantum computer, and the second \ref{Subsec: Noisy Simulation} will be the simulation subject to a noise model.

\subsection{Ideal Quantum Computer}
\label{Subsec: Ideal Simulation}

We will consider a perfect Quantum computer without any noise for the first experiment. The purpose of this experiment is two-fold: firstly, to show that the algorithm works in an ideal setting. Secondly, it will be a fair comparison when we analyze the case with noise, which is of main importance for running Variational Quantum Algorithms in NISQ devices.

\begin{figure}[H]
    \centering
    \includegraphics[width=0.5\textwidth]{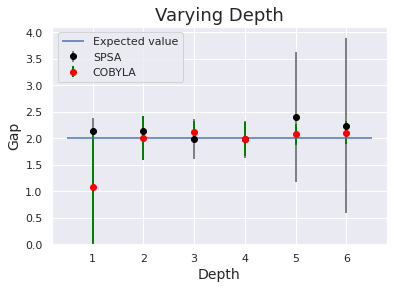}
    \caption{A plot analyzing the depth needed to solve the problem for a two-qubit Hamiltonian using a Hardware Efficient Ansatz, where we took the statistics of 50 runs of the algorithm with random initialization and finite number of shots. Comparing between COBYLA and SPSA optimizers. Analysis of the Optimizer performance, where we used $c=0.7$ and took the average of the last 25 $\lambda_k$ for the final $\lambda_k$ value on the SPSA. As the depth of the circuit increases, the algorithm converges to the solution that is obtained through direct diagonalization.}
    \label{fig:Hardware Efficient Depth Analysis}
\end{figure}

\begin{figure}[H]
    \centering
    \includegraphics[width=0.5\textwidth]{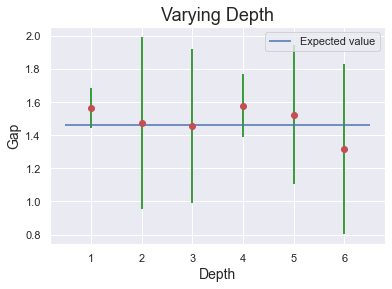}
    \caption{A plot analyzing the depth needed to solve the problem for a five qubit Hamiltonian using a Hardware Efficient Ansatz, where we took the statistics of 50 runs of the algorithm with random initialization and finite number of shots. We choose only to use the COBYLA optimizer in order to show that our algorithm works for a higher number of qubits.}
    \label{fig:Hardware Efficient Depth Analysis 5}
\end{figure}

In order to estimate the depth needed to simulate the Hamiltonian we run the algorithm changing the depth of the ansatz and estimate the gap, this is shown on FIGURES \ref{fig:Hardware Efficient Depth Analysis} and \ref{fig:Hardware Efficient Depth Analysis 5}, for 2 and 5 qubits respectively. For the case of 5 qubits, we choose to only use the COBYLA optimizer, because we have observed that it worked better for the case of 2 qubits.

\small

\small

According to the figures, we observe that for the SPSA optimizer the increase in depth can also lead to higher variance on the samples, this is expected since increasing the depth of the ansatz also leads to a linear increment in the number of parameters, this shows that for this instance the gradient-free optimizer works better for this kind of problem.

In addition, we also see that a depth of 3 is needed to find a reasonable solution for both optimizers to solve the BCS problem for both cases, because it is within 1 standard deviation of the statistical sample and is the smallest depth that satisfies this condition.

\begin{figure}[H]
    \centering
    \includegraphics[width=0.5\textwidth]{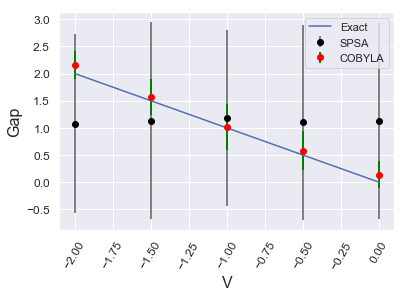}
    \caption{Measuring the gap varying V for the two-qubit Hamiltonian, where we used $c=0.7$ and took the average of the last 25 $\lambda_k$ for the final $\lambda_k$ value for the SPSA optimizer, and we took the statistics of 10 runs of the algorithm with random initialization.}
    \label{fig: V analysis}
\end{figure}

After defining the depth needed to solve the problem, we benchmark the problem of finding the $n=1$ gap of the two-qubit Hamiltonian by changing the coupling parameter $V$, with parameters $\epsilon = \epsilon_1 = \epsilon_2 = 3$. This is done by calculating the first and second excited states of the Hamiltonian. Both optimizers are able to solve the gap problem within 1 standard deviation, which is demonstrated on FIGURE \ref{fig: V analysis}.

\small

However, only the COBYLA optimizer is able to show the expected trend on the mean value. This is due to the need to tune the hyperparameters for each case of the SPSA optimizer.

\begin{figure}[H]
    \centering
    \includegraphics[width=0.5\textwidth]{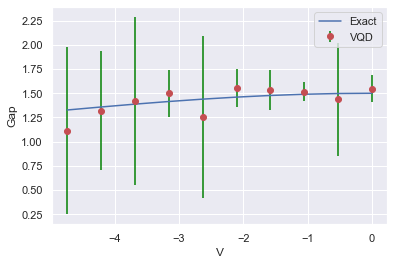}
    \caption{Measuring the gap varying V for the five qubit Hamiltonian using the depth 3, where we took the statistics of 20 runs of the algorithm with random initialization.}
    \label{fig: V analysis 5}
\end{figure}

For the case of the five qubit Hamiltonian, we change the coupling parameter $V$, with $\epsilon_1 = \epsilon_2 = \epsilon_3 = \epsilon_5 = 3$ and $\epsilon_4 = 4$, and estimate the $n=0$ gap, which is done by calculating the gap between the ground state and the first excited state of the Hamiltonian. This is represented in FIGURE \ref{fig: V analysis 5}.

\small

\subsection{Noisy Quantum Computer}
\label{Subsec: Noisy Simulation}

In this work we choose the \textit{Thermal Relaxation Error} \cite{Georgopoulos21Noise} as our noise model which represents the thermalization of the qubit towards the equilibrium state at the temperature of the environment. This is an irreversible process that is governed by the gate time $T_g$, relaxation times $T_1$, $T_2$, and device temperature, $T$, which all can be obtained during the device calibration.

Since the device is kept at a temperature very close to absolute zero, it is safe to assume that $T \approx 0$. This assumption implies that there is no excited state population, which is given by the equation:

\begin{equation}
    \label{Eq: Excited state population}
    p_e = \bigg( 1 + \exp \big( \frac{2hf}{k_B T} \big) \bigg)^{-1}
\end{equation}

Where $T$ is the device's Temperature, $f$ is the qubit frequency, $k_B$ is Boltzmann's constant, and $h$ is Planck's constant. Under this assumption, we have two sources of noise, dephasing and reset to $\ket{0}$.

The relaxation error rates are defined as $\epsilon_{T_1} = e^{-T_g/T_1}$ and $\epsilon_{T_2} = e^{-T_g/T_2}$ and the reset probability is defined as $p_{\mathrm{reset}} = 1 - \epsilon_{T_1}$.

There are two behaviors for this kind of noise: $T_2 \leq T_1$, and $T_2 > T_1$.

For the case, $T_2 \leq T_1$ we can represent the Thermal Relaxation Error as a probabilistic mixture of reset operations and unitary errors. 

The probability of a dephasing happening is given by $ p_{Z} = \frac{1 - p_{\mathrm{reset}}}{2} \bigg( 1 - \frac{\epsilon_{T_2}}{\epsilon_{T_1}} \bigg)$. The probability of a reset to $\ket{0}$ operation occurring is given by $p_{r_{\ket{0}}} = (1 - p_e) p_{\mathrm{reset}}$. The probability of no noise occurring is given by $p_I = 1 - p_{Z} - p_{\mathrm{reset}}$.

Under these conditions, the error model has a Kraus representation as:

\begin{equation}
\begin{split}
    K_I = \sqrt{p_I} I \\
    K_Z = \sqrt{p_Z} Z \\
    K_{\mathrm{reset}} = \sqrt p_{\mathrm{reset}} \ket{i} \bra{0}
\end{split}
\end{equation}

The relaxation noise is represented as the following channel:

\begin{equation}
    \rho \mapsto \mathcal{N}(\rho) = \sum_j K_j \rho K_j^\dagger
\end{equation}

For the case $T_2 > T_1$, the error can be represented by a Choi-Matrix representation. Considering $T \approx 0$, the Choi-Matrix of the Thermal relaxation noise is given by \cite{blank2020quantum}:

\begin{equation}
    \begin{pmatrix}
    1 & 0 & 0 & \epsilon_{T_2} \\
    0 & 0 & 0 & 0 \\
    0 & 0 & p_{\mathrm{reset}} & 0 \\
    \epsilon_{T_2} & 0 & 0 & 1 - p_{\mathrm{reset}} \\
    \end{pmatrix}
\end{equation}

Using the Choi-Matrix representation, it is possible to find the Kraus operators and consequently find the channel for the noise.

It is important to evaluate if the Variational Quantum Algorithm that we are using is resilient to noise in order to run in NISQ devices. Thus, we follow the same methodology as the Ideal Quantum Computer section  \ref{Subsec: Ideal Simulation} under the presence of noise.

We first vary the ansatz depth in order to find the number of layers on the ansatz for the case with two qubits, and we see a similar trend when comparing the analysis with and without noise, however the SPSA optimizer seems to behave not as well as the COBYLA optimizer which shows convergence when the depth increases. This shows that just as without noise, we have a good convergence when the depth of the ansatz is equal to 3, just like the case without noise.

\small
\begin{figure}[H]
    \centering
    \includegraphics[width=0.5\textwidth]{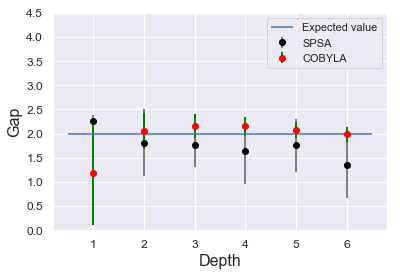}
    \caption{A plot analyzing the depth needed to solve the problem for a two qubit Hamiltonian using a Hardware Efficient Ansatz, where we took the statistics of 10 runs of the algorithm with random initialization and compare between COBYLA and SPSA optimizers. Analysis of the Optimizer performance, where we used $c=0.7$ and took the average of the last 25 $\lambda_k$ for the final $\lambda_k$ value on the SPSA.}
    \label{fig:Hardware Efficient Depth Analysis Noise}
\end{figure}

\begin{figure}[H]
    \centering
    \includegraphics[width=0.5\textwidth]{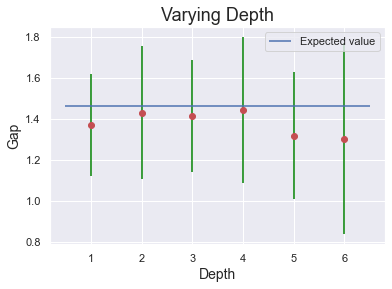}
    \caption{A plot analyzing the depth needed to solve the problem for a five qubit Hamiltonian using a Hardware Efficient Ansatz, where we took the statistics of 50 runs of the algorithm with random initialization for the COBYLA optimizer. The results are adjusted by a constant factor due to the presence of noise.}
    \label{fig:Hardware Efficient Depth Analysis Noise 5}
\end{figure}

For the case with five qubits, we use only the COBYLA optimizer in order to show that our method works for more than two qubits, FIGURE \ref{fig:Hardware Efficient Depth Analysis Noise 5}. The noise leads to an offset which has been removed. We see that, it has a similar behavior than the case without noise, thus choosing the depth of the ansatz equals to 3 works for the five qubit case.

\small

After defining the depth, we implement for the same case, varying $V$ on the Hamiltonian and using $\epsilon = \epsilon_1 = \epsilon_2 = 3$. We observe that for this case, both COBYLA and SPSA optimizers have demonstrated a linear trend, which is expected when the coupling parameter $V$ is changed. 

Even though we used $\epsilon = \epsilon_1 = \epsilon_2$, we could choose $\epsilon_1 \neq \epsilon_2$ without any loss to our algorithm, these values were chosen because it would demonstrate a linear trend on the BCS gap for the case of two qubits, other conditions would lead to non-linear relation when varying the coupling parameter.

\small
\begin{figure}
    \centering
    \includegraphics[width=0.5\textwidth]{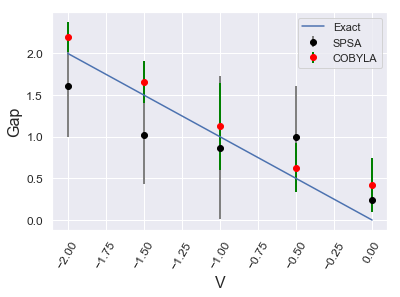}
    \caption{ Measuring the gap varying the coupling constant, $V$, for the two qubit Hamiltonian. We used $c=0.7$ and took the average of the last 25 $\lambda_k$ for the final $\lambda_k$ value on the SPSA.}
    \label{fig:Var V Noise}
\end{figure}

\begin{figure}
    \centering
    \includegraphics[width=0.5\textwidth]{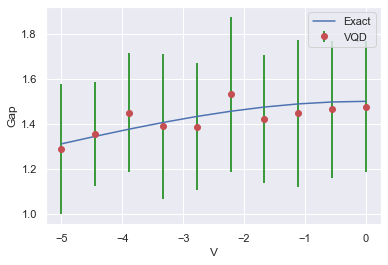}
    \caption{Measuring the gap varying the coupling constant, $V$, for the two qubit Hamiltonian. We use the COBYLA Optimizer performance, to show that our method works with a high qubit number. The results are adjusted by a constant factor due to the presence of noise.}
    \label{fig:Var V Noise 5}
\end{figure}

Now we use the depth of 3, for the case varying the coupling constant, $V$, with $\epsilon_1 = \epsilon_2 = \epsilon_3 = \epsilon_5 = 3$ and $\epsilon_4 = 4$, and estimate the $n=0$ gap. This is represented in FIGURE \ref{fig:Var V Noise 5}.

\small

\section{Conclusion}
\label{Sec: Conclusion}
In this paper we have explored solving the BCS Hamiltonian through the Wu et al. \cite{wu2002BCS2QUBIT} mapping using Near-Term Quantum Computers through a Variational Quantum Algorithm. 

The algorithm that was used is called Variational Quantum Deflation (VQD) and we were able to obtain the $n=1$ gap of a two qubit BCS Hamiltonian within 1 standard deviation considering random initialization, showing that the algorithm works and is not highly dependent of the initialization of the random parameters. We also obtain the $n=0$ gap of a five qubit BCS Hamiltonian, which shows that our methods scale for higher qubit instances.

We also analyzed both gradient and gradient-free optimizers in order to search for variational parameters that solve our task. For the gradient-based optimizer we chose Simultaneous Perturbation Stochastic Approximation (SPSA) and for the gradient-free we chose constraint functions by linear interpolation (COBYLA). Showing that even in the presence of Thermal Relaxation noise the COBYLA optimizer have a good convergence for solving the BCS gap problem, even though it is assumed to be a bad optimizer when the cost function has stochastic noise.

This work shows promising ways to solve the BCS Hamiltonian in near-term quantum computers, which could be extended for an arbitrary number of cooper pairs. For future works, we aim to explore an ansatz that is tailor-made for the Hamiltonian in order to avoid the Barren Plateaus problem and have a better optimization convergence. Another direction is to analyze in more depth the objective function landscape of the VQD algorithm and see if the barren plateau problem is present in this Variational Quantum Algorithm.

\ack
The authors would like to thank João Ribeiro and Filipe Melo for useful comments on this manuscript. This study was financed in part by the Coordena\c c\~ao de Aperfei\c coamento de Pessoal de N\'ivel Superior – Brasil (CAPES) – Finance Code 001, and by the Conselho Nacional de Desenvolvimento Cient\'ifico e Tecnol\'ogico (CNPq). We acknowledge the use of IBM Quantum services for this work. The views expressed are those of the authors, and do not reflect the official policy or position of IBM or the IBM Quantum team. This manuscript used Quantikz \cite{Quantikz} for drawing the circuits and Qiskit \cite{Qiskit} for its simulations. All source code is available on GitHub: \href{https://github.com/nahumsa/volta}{https://github.com/nahumsa/volta}.

\bibliography{Bibliography.bib}

\appendix
\section*{Appendix}
\section{Classical Optimizer}
\label{Appendix: Classical Optimizer}
There are plenty of optimization algorithms that are used on Variational Quantum Algorithms. In this paper, we will focus on two algorithms for optimization: The Simultaneous Perturbation Stochastic Approximation (SPSA) and Constrained Optimization BY Linear Approximations (COBYLA). 

The problem of barren plateaus \cite{mcclean2018barren} is present in gradient-based optimizers, which means that the loss function gradient becomes exponentially smaller making it harder to minimize the cost function. The barren plateaus problem can be induced by various factors such as high ansatz expressiveness \cite{holmes2021expBP} or even presence of noise \cite{wang2020noiseBP}.

Initially, the presence of barren plateaus was found only on gradient-based optimizers, but recently there is evidence that it affects gradient-free optimizers as well \cite{Arrasmith2020GradientFreeBP}.

Therefore, in order to examine the performance of classical optimizers, it is fair to compare gradient-free and gradient based optimizers. We thus choose COBYLA as the gradient-free optimizer and SPSA as the gradient-based optimizer.

\subsection{SPSA}
\label{Subsec: SPSA}
The Simultaneous Perturbation Stochastic Approximation (SPSA) \cite{spall1992SPSA} is an algorithm that is robust  with respect to stochastic perturbations of the function with which it will be minimized and has been widely used in variational quantum algorithms that deal with NISQ devices \cite{kandala2017hardware}. 

The SPSA algorithm works with only two evaluations using a symmetrical Bernoulli distribution $\Delta_i$ and two zero converging sequences, $c_i$ and $a_i$. Since we want to minimize the cost function of the VQD algorithm (eq. \ref{Eq: Cost Function VQD}) we must take the gradient $g(\lambda_k)$ this will be done by sampling the cost function $F(\lambda_k)$, since we only take the sample of the gradient and do not calculate it analytically we name it $\hat{g}(\lambda_k)$ and we obtain it using the formula:

\begin{equation}
    \label{Eq: Gradient SPSA}
    \hat{g}(\lambda_k^{(i)}) = \frac{F(\lambda_k^{(i)} + c_n \Delta_i) - F(\lambda_k^{(i)} - c_i \Delta_i)}{2c_i \Delta_i}
\end{equation}where $\lambda_k^{(i)}$ is the $\lambda_k$ parameter in the $i^{\text{th}}$ iteration of the optimization algorithm.

After evaluating the gradient, the algorithm updates the parameters using the following rule:

\begin{equation}
    \label{Eq: Update rule}
    \lambda_k^{(i+1)} = \lambda_k^{(i)} - a_i \hat{g}(\lambda_k^{(i)})
\end{equation}

It has been proven \cite{spall1992SPSA, maryak2001SPSAConv} that this algorithm can converge assuming stochastic fluctuations in the cost function evaluation. The sequences $a_i$ and $c_i$ are chosen as:
\begin{equation}
    \label{Eq: a_i and c_i sequences}
    \begin{split}
        c_i = \frac{c}{i^\gamma} \\ a_i = \frac{a}{i^\alpha}
    \end{split}
\end{equation}

where the values are optimally chosen \cite{spall1998SPSAParmaters} as $\{ \alpha , \gamma \} = \{ 0.602, 0.101 \}$. The value $a$ controls how large is the update of the parameters and $c$ controls the gradient update, thus if there are large statistical fluctuations on the cost function we must choose a large $c$ in order to the gradient evaluation to be accurate, according to experiments explained in section \ref{Sec:Results} we chose $c=0.7$. 

In addition to that we use a calibration method as in \cite{kandala2017hardware} in order to adjust the $a$ parameter, which starts at $a= 2 \pi / 10$, choosing different directions of $\Delta_i$, the inverse formula used to calibrate $a$ is:

\begin{equation}
    \label{Eq: a calibration}
    a = \frac{2 \pi}{5} \frac{c}{\big\langle F(\lambda_k^{(1)} + c_n ) - F(\lambda_k^{(1)} - c_i) \big\rangle_{\Delta_1} }
\end{equation}

Finally, in order to get the optimized $\lambda_k$ values, we follow \cite{kandala2017hardware} and average over the last 25 iterations in order to suppress statistical fluctuations on parameter values.

\subsection{COBYLA}
\label{Subsec: COBYLA}

Based on a direct search algorithm {\it that models the objective and constraint functions by linear interpolation} \cite{MJDPowell1994} COBYLA is a classical gradient-free optimizer \cite{MJDPowell1998}. It is based on an algorithm that is iterative, in which each iteration produces, by interpolation at the vertices of a simplex, a linear approximation of the objective and constraint functions. The iterative change of values is restricted by a trust region, $\rho$, to assure that the problem has a finite solution. 

It is expected that this kind of optimizer works well when stochastic perturbations are minimal, because it supposes that function evaluation is exact. This could, anyhow, lead to problems when working with real quantum computers that are subject to low decoherence times. However, there is no proof in the literature that this algorithm will not converge with stochastic perturbations, thus we aim to test this case to see if it converges with the presence of noise.

For the trust region, the user needs to choose an initial ($\rho_{\text{init}}$) value that represents the exploration on the objective function landscape, which decreases by half after each iteration in order to find the minimum of the objective function subjected to the given constraints. For our optimization procedure, we choose $\rho_{\text{init}}=1$.

\section{Measuring the overlap of two states on a quantum computer}
\label{Appendix: Swap Test}

In order to use the Variational Quantum Deflation algorithm, we require measuring the overlap between two quantum states. This can be done by three techniques: SWAP test, Destructive SWAP test, and Amplitude Transition. In this section, we will explain each method and comment its advantages and disadvantages.

SWAP test is an algorithm that gives as a result the state overlap between two quantum states $\rho$ and $\sigma$. There are two proposals for a SWAP test in the literature which are equivalent to each other. The original SWAP test \cite{buhrman2001SWAPt} that uses a controlled-SWAP operation and an auxiliary qubit to measure the overlap, and the Destructive SWAP test \cite{cincio2018DSWAP} which only uses first neighbor connectivity, CNOT and 1 qubit gates to measure the overlap.

\begin{figure}[H]
    \centering
    \includegraphics{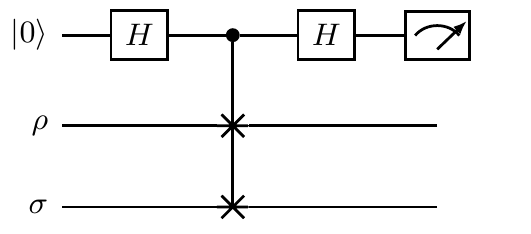}
    \caption{Circuit representation of the SWAP test for states $\rho$ and $\sigma$, this is done using controlled-SWAP gate and 1 qubit gates.}
    \label{fig: SWAP Test}
\end{figure}

The SWAP test is depicted on FIGURE \ref{fig: SWAP Test} and consists of the application of Hadamard and Controlled-SWAP gates, and it is measured using the auxiliary qubit. The probability on measuring the $\ket{0}$ state encodes the overlap between $\rho$ and $\sigma$

\begin{equation}
    \label{Eq: SWAP test outcome}
    |\langle \rho | \sigma \rangle|^2 = 2 \ \bigg( P(0) - \frac{1}{2} \bigg)
\end{equation}

where $P(0)$ is the probability of the outcome on the auxiliary qubit being 0.


\small

This is a solid approach for learning the overlap between two quantum states, however using a controlled-SWAP gate demands high coherence times and high connectivity between qubits, those two requirements are not available on current quantum hardware. For NISQ devices we must use an algorithm that is suited for the device's restrictions on both connectivity and gate errors, the algorithm that solve both those restrictions in certain conditions is the Destructive SWAP test.

\begin{figure}[H]
    \centering
    \includegraphics{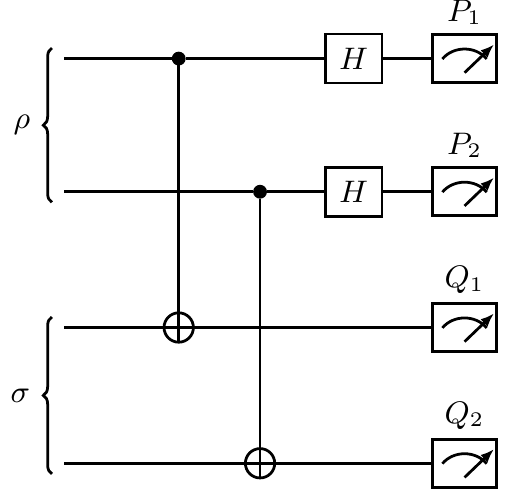}
    \caption{Destructive SWAP test for states $\rho$ and $\sigma$, after the measurement on the Bell basis, it is needed to do a post-processing classically.}
    \label{fig: Destructive Swap Test}
\end{figure}

The Destructive SWAP test is represented on FIGURE \ref{fig: Destructive Swap Test} which consists of CNOT and Hadamard Gates and no auxiliary qubit. This algorithm has depth $O(1)$ which makes it suited for NISQ devices and has demonstrated its superiority against the original SWAP Test \cite{cincio2018DSWAP}. 

In order to obtain the state overlap, one ought to do measurements in the Bell basis for the corresponding qubit in each state. Those measurements can be achieved with a classical post-processing step that scales linearly with qubit size. This post-processing can be interpreted as  the expectation value of a controlled-Z observable.


\small

It is important to analyze the topology needed for these algorithms to be used in real quantum hardware that are available through the cloud. In order to implement the SWAP test \cite{buhrman2001SWAPt} in real hardware, even though the cost of a swap gate is high depending on the hardware topology, we could implement it without making various re-routing of qubits. This topology is supported by the IonQ device provided by Amazon Braket \cite{Braket}.

\begin{figure}[H]
    \centering
    \begin{tikzpicture}[main/.style = {draw, circle}] 
    \node[main] (1) {$p_1$}; 
    \node[main] (2) [right of=1] {$p_2$};
    \node[main] (3) [right of=2] {$p_3$}; 
    \node[main] (9) [above of=2] {$a_1$}; 
    \node[main] (4) [below of=1] {$q_1$};
    \node[main] (5) [right of=4] {$q_2$}; 
    \node[main] (6) [right of=5] {$q_3$};
    \node[]  (7) [right of=3] {$\dots$};
    \node[]  (8) [right of=6] {$\dots$};
    
    \draw (1) -- (2);
    \draw (1) -- (4);
    \draw (2) -- (3);
    \draw (2) -- (5);
    \draw (3) -- (6);
    \draw (3) -- (7);
    \draw (4) -- (5);
    \draw (5) -- (6);
    \draw (6) -- (8);
    \draw (9) -- (1);
    \draw (9) -- (2);
    \draw (9) -- (3);
    \draw (9) -- (4);
    \draw (9) -- (5);
    \draw (9) -- (6);
    \end{tikzpicture} 
    \caption{Hardware topology needed for the SWAP Test.}
    \label{fig: Hardware topology SWAP}
\end{figure}
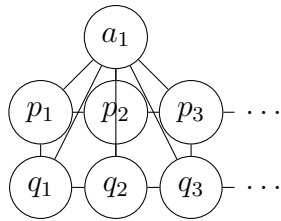

\begin{figure}[H]
    \centering
    \begin{tikzpicture}[main/.style = {draw, circle}] 
    \node[main] (1) {$p_1$}; 
    \node[main] (2) [right of=1] {$p_2$};
    \node[main] (3) [right of=2] {$p_3$}; 
    \node[main] (4) [below of=1] {$q_1$};
    \node[main] (5) [right of=4] {$q_2$}; 
    \node[main] (6) [right of=5] {$q_3$};
    \node[]  (7) [right of=3] {$\dots$};
    \node[]  (8) [right of=6] {$\dots$};
    
    \draw (1) -- (2);
    \draw (1) -- (4);
    \draw (2) -- (3);
    \draw (2) -- (5);
    \draw (3) -- (6);
    \draw (3) -- (7);
    \draw (4) -- (5);
    \draw (5) -- (6);
    \draw (6) -- (8);
    \end{tikzpicture} 
    \caption{Hardware topology needed for the Destructive Swap Test.}
    \label{fig: Hardware topology DSWAP}
\end{figure}
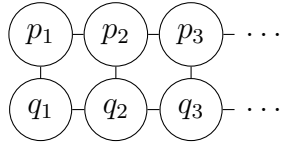

In order to implement the Destructive SWAP test in real hardware, it is necessary to have a specific hardware topology that is represented in FIGURE \ref{fig: Hardware topology DSWAP}. Where the first row of the graph encodes one quantum state and the second encodes the second quantum state, and near-neighbor connection between them is used for the Destructive SWAP test. There are actual devices that support this kind of topology, for instance there is the Melbourne chip from IBMQ \cite{IBMQ} and the IonQ device provided by Amazon Braket \cite{Braket}.

\begin{figure}[H]
    \centering
    \includegraphics{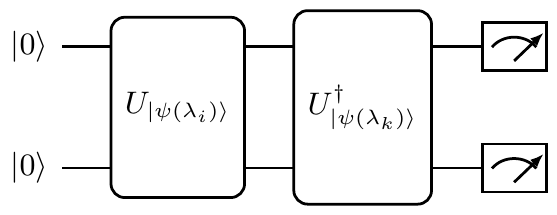}
    \caption{Circuit representation of the transition amplitude method \cite{Havlek2019} for measuring the overlap between $\ket{\psi(\lambda_i)}$ and $\ket{\psi(\lambda_k)}$.}
    \label{fig: Amplitude Transition Test}
\end{figure}

It is possible to measure the overlap between two states using the transition amplitude between the two states \cite{Havlek2019}: $\big| \langle \psi(\lambda_k) | \psi(\lambda_i) \rangle \big|^2 = \big| \langle 0 | U^\dagger_{\psi(\lambda_k)}  U_{\psi(\lambda_i)} | 0 \rangle \big|^2$, where $U_{\psi(\lambda_i)}$ is the state generated by the variational ansatz. The overlapping is obtained by measuring the frequency of obtaining the $\ket{0}$ state. This approach leads to no restriction for the hardware. However, it doubles the depth of the circuit, which exceed the depth that is suited for NISQ devices. This procedure is represented on FIGURE \ref{fig: Amplitude Transition Test}. 


\small

To summarize, if the topology of the hardware supports the FIGURE \ref{fig: Hardware topology SWAP} topology, using the SWAP test can be an advantage because you would be adding only 3 gates to your circuit, thus not growing the depth to the point that will make it impossible to run on NISQ devices. If the hardware supports FIGURE \ref{fig: Hardware topology DSWAP}, using the destructive SWAP test could be an advantage because you would only need a constant number of gates added in the end of the circuit, since the overlap can be easily calculated using classical post-processing. It may appear that the transition amplitude method is a superior method than the other methods because it doesn't imply any restrictions of the topology, however there is a huge downside that comes with doubling the depth of the circuit, which can make it worse than the Destructive SWAP test, and the SWAP test in some hardware topologies.

\end{document}